\documentclass[sigconf]{acmart}
\usepackage{algorithm}
\usepackage{placeins}
\usepackage{algorithmic}
\usepackage{float}

\AtBeginDocument{%
  }

\acmConference[FL@FM-TheWebConf'25]{International Workshop on Federated Foundation Models for the Web 2025
}{April 29,
  2025}{Sydney, Australia}
\acmBooktitle{International Workshop on Federated Foundation Models for the Web 2025,  April 25, 2025, Sydney, Australia}

\begin{document}

\newcommand{\methodname}{{\tt{FedCDC}}}
\title{FedCDC: A Collaborative Framework for Data Consumers in Federated Learning Market}


\author{Zhuan Shi}
\authornote{Both authors contributed equally to this research.}
\email{zhuan.shi@epfl.ch}
\affiliation{
  \institution{École Polytechnique Fédérale de Lausanne (EPFL)}
  \city{Lausanne}
  \country{Switzerland}
}

\author{Patrick Ohl}
\authornotemark[1]
\email{patrick.ohl@epfl.ch}
\affiliation{%
  \institution{École Polytechnique Fédérale de Lausanne (EPFL)}
  \city{Lausanne}
  \country{Switzerland}
}

\author{Boi Faltings}
\email{boi.faltings@epfl.ch}
\affiliation{
  \institution{École Polytechnique Fédérale de Lausanne (EPFL)}
  \city{Lausanne}
  \country{Switzerland}
}

\renewcommand{\shortauthors}{Shi and Ohl et al.}

\begin{abstract}
Federated learning (FL) allows machine learning models to be trained on distributed datasets without directly accessing local data. In FL markets, numerous Data Consumers compete to recruit Data Owners for their respective training tasks, but budget constraints and competition can prevent them from securing sufficient data. While existing solutions focus on optimizing one-to-one matching between Data Owners and Data Consumers, we propose \methodname{}, a novel framework that facilitates collaborative recruitment and training for Data Consumers with similar tasks. Specifically, \methodname{} detects shared subtasks among multiple Data Consumers and coordinates the joint training of submodels specialized for these subtasks. Then, through ensemble distillation, these submodels are integrated into each Data Consumer’s global model. Experimental evaluations on three benchmark datasets demonstrate that restricting Data Consumers’ access to Data Owners significantly degrades model performance; however, by incorporating \methodname{}, this performance loss is effectively mitigated, resulting in substantial accuracy gains for all participating Data Consumers.
\end{abstract}

\ccsdesc[500]{Computing methodologies~Machine learning}
\ccsdesc[500]{Computing methodologies~Distributed artificial intelligence}

\keywords{Federated Learning, Ensemble Distillation, Collaborative Training}


\maketitle

\section{Introduction}

Federated Learning (FL) \cite{mcmahan2017communication} enables model training on distributed datasets without revealing the underlying local data. By training models locally on each device and then aggregating their updates on a central server, FL preserves privacy while leveraging diverse data sources. When a Data Consumer (DC) aims to build a model using sensitive data from multiple devices, traditional machine learning approaches are typically infeasible due to privacy constraints. Instead, the DC employs FL and compensates each Data Owner (DO)—e.g., through monetary payments—for participating in the training process.

To maximize model performance, a DC must recruit a wide range of high-quality DOs. However, DOs incur computation and communication costs during FL training and therefore need adequate incentives to participate. Under limited budgets, DCs may be unable to recruit sufficient DOs. Game-theoretic approaches—such as auctions, smart contracts, or core stability—have been used to manage these DC-DO interactions \cite{tu2022incentive, zhang2021incentive, zhou2021truthful, zhan2022survey, le2020cellular, zhang2022onlineauctionbasedincentivemechanism, khan2020federated, ray2022fairness}.

\begin{figure}[h]
\begin{center}
\includegraphics[width=\linewidth]{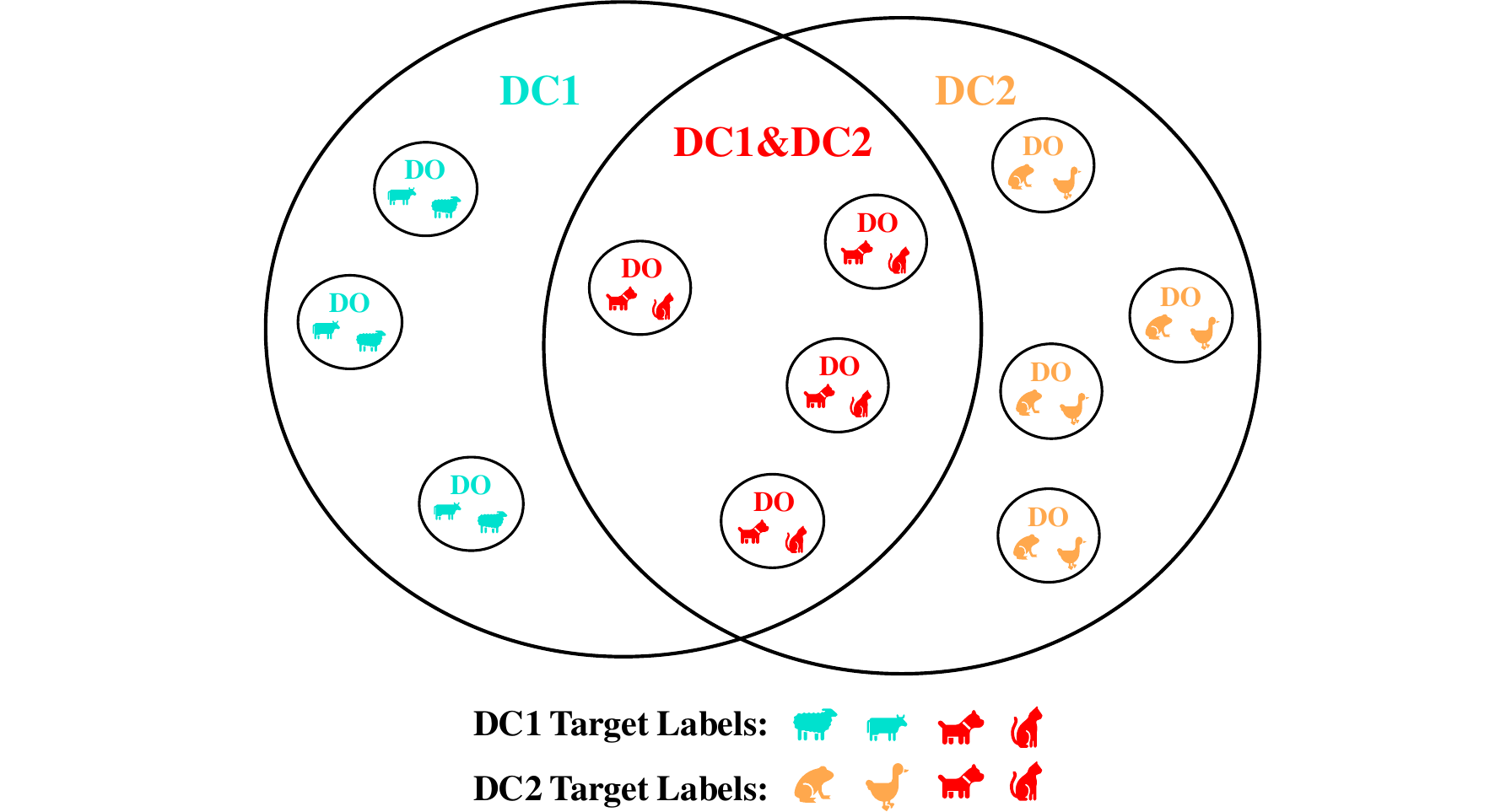}
\end{center}
\caption{Two DCs with overlapping target domains. Both DCs want to classify dogs and cats, so they will both try to recruit the red DOs who hold dog and cat data.}
\label{fig:target_domains}
\Description[Intersecting Target Domains]{The target domains of Data Consumer 1 and Data Consumer 2 are shown in a Venn diagram. Multiple Data Owners lie in different parts of the diagram, depending on the data they hold. Data Owners in the intersection of the target domains are red.}
\end{figure}

\begin{figure}[h]
\begin{center}
\includegraphics[width=0.9\linewidth]{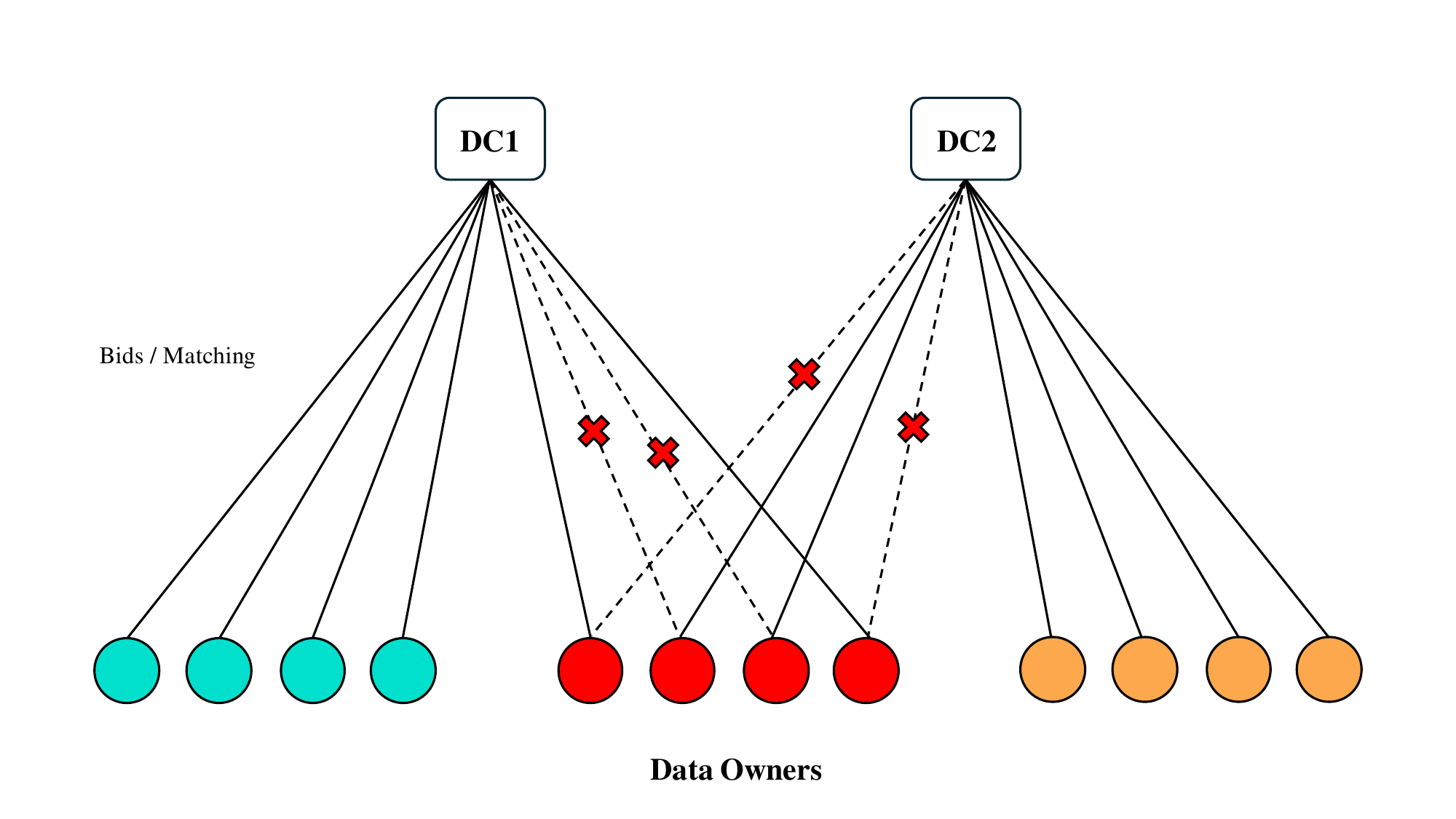}
\end{center}
\caption{A DC-DO matching in a FL market with 2 DCs and 12 DOs. Both DCs try to recruit the red DOs, but due to limited computational resources, a DO can only be recruited by one DC. Therefore, the DCs fail to recruit all the DOs they want.}
\label{fig:do_dc_matching}
\Description[Simple DO-DC Matching]{A diagram depicting a matching of Data Owners to Data Consumers. There is a line from each Data Owner to the Data Consumer it is matched to. If the Data Consumer bid for the Data Owner, but did not manage to recruit it, there is a dashed line. Both Data Consumers bid for the red Data Owners.}
\end{figure}

In practical scenarios, multiple DCs may simultaneously compete for DOs \cite{tang2024stakeholder}, as illustrated in Figure~\ref{fig:target_domains}. Because DOs often run on resource-constrained devices, each DO can train only one model at a time. Consequently, when multiple DCs vie for the same DO, only one can recruit it (see Figure~\ref{fig:do_dc_matching}), restricting other DCs’ training processes. Existing game-theoretic and recruitment strategies in multi-DC settings \cite{tang2023competitive, mai2022automatic} primarily address matching mechanisms between DCs and DOs. However, they do not resolve the fundamental problem of competition for the same DO.

Most current methods view DCs as purely competitive, and while \cite{tang2023competitive} introduces collaborative bidding strategies, it does not circumvent the limitation that a DO can only support one DC at a time. To address this shortcoming, we propose a collaborative framework enabling DCs to share DOs, even when they require the same data. For example, one DC might aim to classify insects, while another classifies volant (flying) animals; both would need DOs that possess data on flying insects. Our approach identifies these shared subtasks and leverages them to foster collaboration among DCs.

Specifically, our FL market platform identifies DCs with overlapping tasks competing for the same DOs. If all participating DCs agree, an auxiliary (artificial) DC focusing on the shared subtask is instantiated. Once trained, this auxiliary model is integrated into each participating DC’s global model through ensemble distillation. By pooling resources for shared DOs, DCs effectively expand their budgets and mitigate the performance loss caused by restricted DO access.

Despite its benefits, this collaboration approach presents two main challenges:

\begin{itemize} 
\item \textbf{Heterogeneous model architectures.} DCs may use vastly different model types and sizes, rendering simple weight-merging methods such as FedAvg inapplicable. \item \textbf{Complex collaboration patterns.} Since DCs are self-interested, they decide on their participation and partners independently. In a large FL market, intricate collaboration structures can emerge as DCs vary in their shared subtasks and interests. 
\end{itemize}

To overcome these challenges, we introduce \methodname{}, a novel collaborative framework for federated learning markets. \methodname{} uses ensemble distillation—a model-fusion strategy that accommodates varying model architectures in heterogeneous data environments \cite{lin2020ensemble, cho2022heterogeneousensembleknowledgetransfer, deng2023hierarchical, wang2023dafkd, park2024overcoming, li2019fedmdheterogenousfederatedlearning}. We formulate the process of alliance creation as a weighted max-clique problem which we solve using \textsc{TSM-MWC} \cite{jiang2018maxsat} to identify the optimal collaboration structure. We applies MaxSAT reasoning to solve the weighted maximum clique problem using a two-stage approach: First, it reduces the search space by pruning infeasible or suboptimal subsets, and then it utilizes refined MaxSAT formulations to efficiently identify the optimal solution.

Our main contributions are as follows:
\begin{itemize} 
\item We analyze how multi-DC competition affects model performance in FL markets.
\item We introduce the concept of alliances to implement DC collaboration, and propose \methodname{}, an ensemble-distillation-based framework that dynamically forms these alliances.
\item Through experiments on three benchmark datasets, we show that restricting DC access to DOs significantly impairs model performance; however, integrating \methodname{} largely mitigates this issue, resulting in notable performance gains for all participating DCs.
\end{itemize}

\section{Preliminaries and Related Work}
\subsection{Federated Learning}
Federated Learning (FL) \cite{mcmahan2017communication, yang2019federated, liu2024recent} is a privacy-preserving, distributed machine learning technique that enables the training of a global model on geographically dispersed datasets without centralizing the data. Instead, the server shares its model with devices holding the data, and these devices perform local training. The locally updated models are then sent back to the server, which aggregates them into a single global model.

The \textsc{FedAvg} framework \cite{mcmahan2017communication} employs standard stochastic gradient descent (SGD) for local training and uses a weighted averaging scheme for aggregation, where the weights are proportional to the size of each local dataset (Equation.\eqref{eq:fedavg_aggregation}). To enhance computational and communication efficiency, \textsc{FedAvg} randomly samples a subset of clients in each training round: \begin{equation} \label{eq:fedavg_aggregation} w_{t+1} \gets \sum_{k=1}^K \frac{|D_k|}{\sum_{k=1}^K |D_k|} w_t^k \end{equation}

Although \textsc{FedAvg} performs well in settings where data and system resources are relatively uniform, it struggles in the presence of system and statistical heterogeneity—namely, when clients vary in their computational and communication capabilities or their local data follow different distributions. Various methods have been proposed to address these challenges. For example, \textsc{FedProx} introduces a proximal term (Equation. \eqref{eq:fedprox}) to the local loss function, constraining local updates to stay close to the global model and thereby mitigating the adverse effects of statistical heterogeneity: \begin{equation} \label{eq:fedprox} \frac{\mu}{2}\lVert w_t^k - w_t \rVert^2 \end{equation}

More sophisticated aggregation techniques, such as \textsc{FedMA} \cite{wang2020federated}, have also been developed to further address heterogeneity. Recently, ensemble distillation has attracted significant attention as an alternative aggregation method \cite{lin2020ensemble, cho2022heterogeneousensembleknowledgetransfer, deng2023hierarchical, wang2023dafkd, park2024overcoming, gong2021ensemble, li2019fedmd}, as it allows for more flexible integration of local models with varying architectures and data distributions.

\subsection{Federated Ensemble Distillation}
Knowledge Distillation \cite{gou2021knowledge, phuong2019towards} can be used to transfer the knowledge of a teacher model to a student model. It works by making the student model mimic the behaviour of the teacher model. Knowledge Distillation therefore requires a dataset that may consist of real data or synthetic data generated by a GAN \cite{micaelli2019zero}. The student then learns by minimizing a loss that is based on a measure of distance between the student and the teacher logits or probability distributions. For the distance function, the KL-Divergence (Equation.\eqref{eq:kl_div}) between the student and teacher probability distributions is often used.
    
\begin{equation}
    D_\text{KL}(P_1, P_2)= \sum_xP_1(x) log\frac{P_1(x)}{P_2(x)}
    \label{eq:kl_div}
\end{equation}

When the distillation dataset is unlabeled, pseudo labels can be produced from the teacher logits, in order to incorporate a cross-entropy loss.

Ensemble Distillation \cite{allenzhu2023understandingensembleknowledgedistillation, wu2022unifiedeffectiveensembleknowledge, vongkulbhisal2019unifyingheterogeneousclassifiersdistillation} is a variant of Knowledge Distillation in which multiple teacher models 
$\{\mathcal{T}_1, ..., \mathcal{T}_{N_T}\}$
collectively transfer their knowledge to a single student model. A simple way to achieve this is by averaging the logits produced by the teachers, thereby integrating insights from heterogeneous teacher models without requiring direct access to their parameters. In Federated Learning (FL), this approach allows different Data Owners (DOs) to train models tailored to their computational resources. Consequently, ensemble distillation in FL has received considerable attention in recent years \cite{lin2020ensemble, cho2022heterogeneousensembleknowledgetransfer, deng2023hierarchical, wang2023dafkd, park2024overcoming}, with a key focus on how best to weight the teacher logits during the averaging process.

Originally, \cite{lin2020ensemble} employed uniform weighting for teacher logits (Equation.\eqref{eq:feddf_weighting}),
\begin{equation} \label{eq:feddf_weighting} z^\mathcal{T} = \frac{1}{N_T} \sum_{i=1}^{N_T} z^{\mathcal{T}_i}, \end{equation} but subsequent work explored more sophisticated weighting schemes based on variance \cite{cho2022heterogeneousensembleknowledgetransfer} or entropy \cite{deng2023hierarchical}. In this paper, we adopt an entropy-based approach.

Notably, \textsc{FedHKT} \cite{deng2023hierarchical} is a hierarchical FL framework that clusters DOs into groups, trains an expert model for each group via \textsc{FedAvg}, and then merges these experts using ensemble distillation. This strategy closely aligns with our objective of training a shared submodel and integrating it into the global Data Consumer (DC) models.

We assume the availability of an unlabeled public dataset $D_{public}$ for the distillation process. In settings where no such dataset exists, zero-shot approaches—such as training a Generative Adversarial Network (GAN) \cite{micaelli2019zero}—can be employed to generate synthetic data for the procedure.

\subsection{FL Markets}
A FL market consists of 
\begin{itemize}
    \item Data Consumers $\{C_1, ..., C_m\}$. A data consumer $C_i=(M_i, L_i, D_{V_i})$ wants to train its model $M_i$ to perform well on its target task. From a data-centric perspective, a target task corresponds to a target domain, on which the DC wants to train its model. In this paper, we focus on image classification tasks, in which a target domain is determined by the classes $L_i$ that $C_i$ wants to classify. In practice, the target domain of $C_i$ is represented by the DC validation set $D_{V_i}$. Since the Data Consumers are indexed, we can refer to them as $\mathcal{C}=\{1, ..., m\}$, i.e. identify them by their indexes.
    \item Data Owners $\{O_1, ..., O_n\}$ who own training sets $\{D_1, ..., D_n\}$ respectively. We will also identify DOs by their indexes $\mathcal{O}=\{1, ..., n\}$
    \item A FL platform with a DO-DC-matching and a payment determination mechanism. This could for example be an auction.
\end{itemize}

The FL platform assigns available DOs to DCs and determines adequate payments to compensate DOs for the incurred training costs. The DCs use their assigned DOs in the FL training process.

The matching and payment determination mechanism is an area of active research. DCs and DOs are self-interested agents whose knowledge about each other is limited due to the privacy constraints in FL scenarios. While DCs want to recruit high-quality DOs to maximize their model performance, DOs try to maximize their reward. Therefore, determining an optimal matching and corresponding payments while respecting incentive compatibility is hard. Research has been conducted to optimize the market mechanism \cite{mai2022automatic}, as well as the strategies of the participants \cite{mai2022automatic, tang2023utility,tang2024agentorientedjointdecisionsupport, tang2024bias}.

\section{Our Approach}

\begin{figure}[h]
\begin{center}
\includegraphics[width=\linewidth]{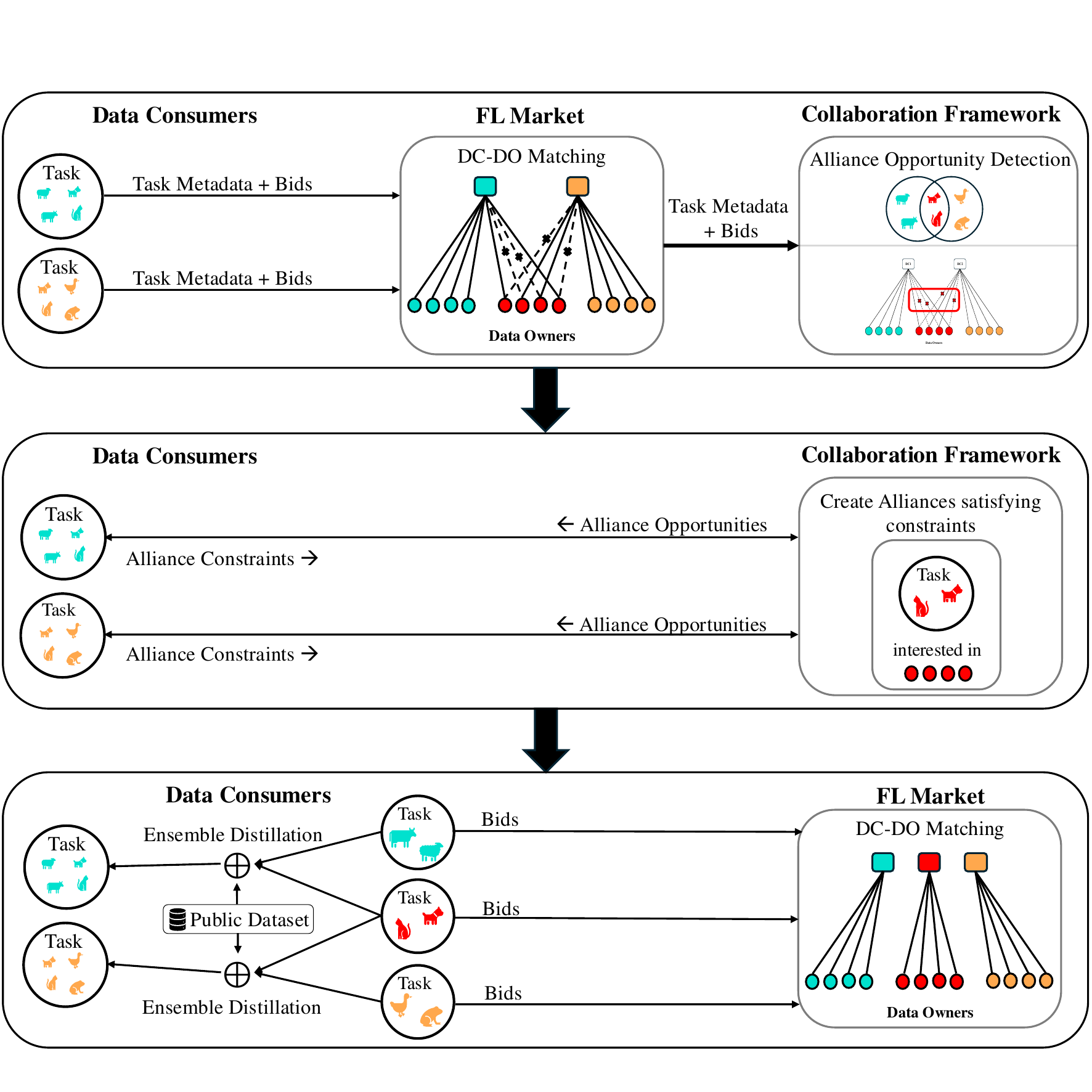}
\end{center}
\vspace{-0.2in}
\caption{An overview of how \methodname{} enables collaboration in a simple setting with 2 Data Consumers. The framework detects that the DCs share the subtask of classifying dogs and cats, so it proposes to create an alliance. Both DCs accept, so an artificial DC is created which will be trained on the shared subtask. Its knowledge is then distilled into the models of the two DCs using ensemble distillation.}
\label{fig:framework}
\Description[FedCDC Flowchart]{General workflow are explained in the caption. Details are explained in the text.}
\end{figure}

Our goal is to increase the access of DCs to high-quality DOs by enabling collaboration between Data Consumers. Different DCs with similar tasks will compete for DOs, since they are interested in similar data. By collaboratively training a model for this common subtask and then integrating it into the different DC models, we can reduce competition and use DOs more efficiently. We therefore propose a framework that allows Data Consumers to dynamically create alliances in order to solve common subtasks. A general overview of the workflow of \methodname{} in a simple setting is given in Figure. \ref{fig:framework}.

\subsection{Alliances}
In order to enable Data Consumer Collaboration, we introduce the concept of alliances.  An alliance $A=(\pi(A), C_A, \delta(A))$ consists of
\begin{itemize}
    \item A set of $n_A$ Data Consumers who participate in the alliance: $\{C_{i_1}, ..., C_{i_{n_A}}\}$. Since the Data Consumers are indexed, we can define the participants as $\pi(A)=\{i_1, ..., i_{n_A}\}$
    \item An artificial data consumer $C_A=(M_A, L_{\pi(A)}, D_{V_A})$ with a model $M_A$ and a task that is shared between all participating Data Consumers. This shared task corresponds to the intersection of the target labels of the participants.
    
    \begin{displaymath}
        L_{\pi(A)}=\bigcap_{i\in\pi(A)} L_i
    \end{displaymath}
        
    Since the validation datasets of the Data Consumers in the alliance may be sensitive, the artificial DC does not have its own validation set. However, the DCs in the alliance can perform decentralized evaluation of the alliance model by using the adequate subset of their validation set. Therefore, the artificial DC effectively has its own validation set.
    \begin{displaymath}
        D_{V_A}=\bigcup_{i\in\pi(A)} \{x\in D_i: label(x) \in L_{\pi(A)}\}
    \end{displaymath}
    \item The set of Data Owners $\delta(A)$ for which the participants in A previously competed.
    \item A budget allocation mechanism $B$. Since the artificial DC does not have its own budget, $\{C_{i_1}, ..., C_{i_{N_A}}\}$ need to allocate parts of their budgets to $C_A$, such that it is able to recruit Data Owners. 
    \begin{equation}
         b(C_A)= \sum_{C_i\in \pi(A)} payment(C_i \rightarrow A)
         \label{eq:alliance_budget}
    \end{equation}

\end{itemize}
Furthermore, introducing the concept of alliances entails the following modifications to Data Consumers:
\begin{itemize}
    \item When a DC is in an alliance, it should not try to recruit DOs that the alliance DC tries to recruit.
    \item The DC needs an aggregation method in order to merge the alliance models with its own model. We propose Ensemble Distillation to achieve this, since the models of different DCs may have different architectures. We call the model that the DC trains on the DOs outside of the alliances target domains $M'_i$. The models are then merged into the global model of the DC.
    \item The DC agents need to decide when they want to participate or stay in an alliance. Furthermore, the budget allocation mechanism of an alliance may require further decision-making, for example, if $B_A$ is an auction.
\end{itemize}

We focus on the design of a collaboration workflow that can be integrated into a large variety of FL Markets, as well as the aggregation method for the DCs. The design of the budget allocation mechanism, as well as the modification of the DC agents are out of the scope of this paper.

\begin{algorithm}[!t]
   \caption{Alliance Creation: \textsc{CreateAlliances}}
   \label{alg:create_alliances}
\begin{algorithmic}
   \STATE {\bf Input:} Data Consumers $\mathcal{C}$, $|\mathcal{C}|=m$; Data Owners $\mathcal{O}$, $|\mathcal{O}|=n$; Bidding history $B^{hist}\in\mathbb{R}_{\geq 0}^{k, m, n}$, where for each matrix $B_j$, each row $B_i$ are the bids of $C_i$ in round $j$; Minimum label overlap $n_{L, min}$; Minimum DO overlap $n_{\delta, min}$
   \STATE
   \FORALL{$i \in \mathcal{C}, o \in \mathcal{O}$}
   \STATE $B^{max}_{i, o} \gets max_{j\in[1, ..., k]} B^{hist}_{j, i, o}$
   \ENDFOR
   \STATE Possible Alliances: $\mathcal{A} \gets \{\}$
   \FORALL{subsets $\pi(A) \subseteq \mathcal{C}$}
        \STATE Common Labels: $L_{\pi(A)}=\bigcap_{C_i\in\pi(A)} L_i$
        \IF{$|L_{\pi(A)}|\geq n_{L, min}$}
        \FORALL{$o \in \mathcal{O}$}
            \STATE $B^{\pi(A)}_o=\prod_{i \in \pi(A)} B^{max}_{i, o}$
        \ENDFOR
            \STATE $\delta(A) \gets\{ o \mid o \in \mathcal{O} : B^{\pi(A)}_o \neq 0 \}$
            \IF{$|\delta(A)| \geq n_{\delta, min}$}
                \STATE Generate unique id $u$
                \STATE $A_u \gets (u, \pi(A), L_\pi(A), \delta(A))$
                \STATE $\mathcal{A} \gets \mathcal{A} \cup \{A_u\}$
            \ENDIF
        \ENDIF
   \ENDFOR
   \STATE Rejected alliance combinations: $\mathcal{U}^- \gets \{\}$
   \FORALL{$i \in \mathcal{C}$}
\STATE $\mathcal{A}_i \gets \{A \in \mathcal{A}:i \in \pi(A)\}$
   \STATE $\mathcal{A}_i^{anon}\gets \{(u, |\pi(A)|, L_\pi(A), \delta(A)) \mid A \in \mathcal{A}_i\}$
   \STATE Send $\mathcal{A}_i^{anon}$ to $C_i$
   \STATE Receive uids of $C_i$'s accepted alliances: $U_i$
   \STATE Receive uids of rejected pairs of alliances: $\mathcal{U}_i^-$
   \FORALL{$A_u \in \mathcal{A}_i$}
    \IF {$u \notin U_i$}
    \STATE $\mathcal{A} \gets \mathcal{A} \setminus \{A_u\}$
    \ENDIF
   \ENDFOR
   \STATE $\mathcal{U}^- \gets \mathcal{U}^- \cup \mathcal{U}_i^-$
   \ENDFOR
    \STATE $V=\{u \mid A_u \in \mathcal{A}\}$
    \STATE $\text{Define } g:V \rightarrow\mathbb{N},    g(u) = |\pi(A_u)|\cdot |L_{\pi(A_u)}| \cdot|\delta(A_u)|$
    \STATE $E=\{(u_1, u_2) \mid A_{u_1}, A_{u_2} \in \mathcal{A} : (u_1, u_2) \notin \mathcal{U}_i^-\}$
   \STATE Create undirected graph $G=(V, E)$ 
   \STATE $U^* = \arg\max_{U \subseteq V} \sum_{u \in U} g(u), \quad \text{s.t. } \forall u, v \in U, (u, v) \in E$
   \STATE $\mathcal{A}_{final} \gets \{\}$
   \FORALL{$u \in U^*$}
    \STATE $D_{V_{A_u}}=\bigcup_{i\in\pi(A_u)} \{x\in D_i: label(x) \in L_{\pi(A_u)}\}$
    \STATE $M_{A_u} \gets ResNet18$ (or another model)
    \STATE $C_{A_u} \gets (M_{A_U}, L_{\pi(A_u)}, D_{V_{A_u}})$
    \STATE $\mathcal{A}_{final} \gets \mathcal{A}_{final} \cup \{(\pi(A_u), C_{A_u}, \delta_{A_u})\}$
   \ENDFOR
    \STATE {\bf Output:} Created Alliances $\mathcal{A}_{final}$
\end{algorithmic}
\end{algorithm}

\subsection{Alliance Creation Procedure}
In order to integrate the concept of alliances into an FL market, the
framework must be able to dynamically detect alliance opportunities and create alliances when the corresponding DCs are willing to participate. In Algorithm. \ref{alg:create_alliances}, we describe this procedure for \methodname{} in detail. We first check for which combinations of DCs the shared subtask is significant enough such that collaboration makes sense. Since \methodname{} is specialized on image classification, this corresponds to the size of the intersection of classes that the DCs want to classify. The minimum necessary overlap size is $n_{L,min}$. We then verify whether the DCs are interested in many of the same DOs.
If they have bid for at least $n_{\delta,min}$ of the same DOs in the last $k$ rounds, we classify them as competing DCs which are therefore eligible for collaboration. Once \methodname{} has detected all alliance opportunities, it sends the metadata of each possible alliance to the corresponding DCs. This metadata consists of the number DCs in the alliance, the number of DOs that the alliance would try to recruit, as well as the labels, in which the alliance would be interested. Based on this information, each DC decides which alliances it wants to participate in. Furthermore, it can specify which pairs of alliances it does not want to simultaneously participate in. This is crucial, because multiple proposed alliances can be very similar and it may therefore be counterproductive to participate in all of them. Once the framework has received the responses from all DCs, it knows which alliances can be created and which pairs of alliances can not be created simultaneously. It then finds a combination of alliances which satisfies the constraints given by the DCs and which maximizes the function given in Equation. \eqref{eq:objective_function}.
\begin{equation}
    \label{eq:objective_function}
    v(\{A_1, ..., A_z\}) = \sum_{i=1}^z|\pi(A_i)|\cdot |L_{\pi(A_i)}| \cdot|\delta(A_i)|
\end{equation}

We choose this objective function, because more DCs, a bigger task overlap and more DOs of interest increase the value of an alliance. In order to obtain a solution to this constrained maximization problem, we can solve the weighted max-clique problem on the following graph $G=(V, E)$. 

\begin{itemize}
    \item Every possible alliance $A$ corresponds to a node with weight $g(A)$.
    \begin{displaymath}
    g(A) = |\pi(A)|\cdot |L_{\pi(A)}| \cdot|\delta(A)|
\end{displaymath}
    \item There exists an edge between the nodes of all pairs of alliances that are not reported as conflicting by any DC.
\end{itemize}

The solution to this weighted max-clique problem will be a set of alliances $\mathcal{A}$, such that
\begin{itemize}
    \item For any alliance $A \in \mathcal{A}$, all participants $\pi(A)$ are willing to participate in the alliance, because otherwise, the corresponding node would not have been added to the graph.
    \item $\mathcal{A}$ does not contain any pair of alliances that a DC has marked as conflicting, since a clique has to be fully connected, so an edge between the nodes of any two alliances has to exist.
    \item $\mathcal{A}$ maximizes our objective function, since it maximizes
     \begin{displaymath}
        \sum_{v \in V}g(v) = \sum_{A \in \mathcal{A}}|\pi(A)|\cdot |L_{\pi(A)}| \cdot|\delta(A)|
    \end{displaymath}
\end{itemize}

\begin{figure}[h]
\begin{center}
\includegraphics[width=\linewidth]{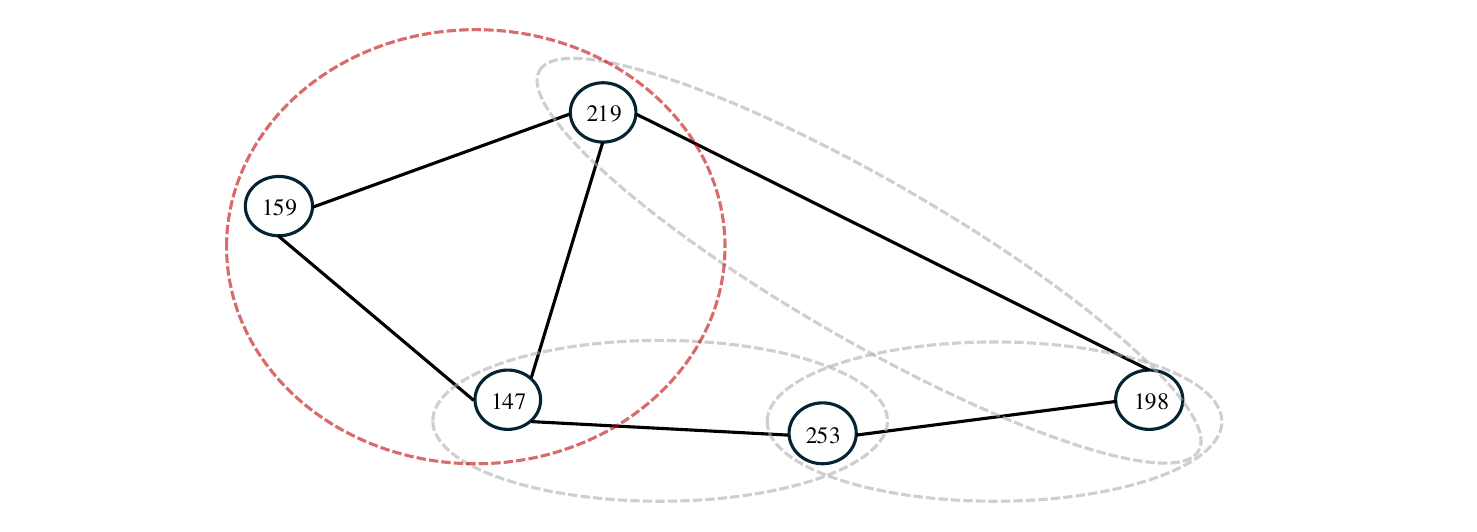}
\end{center}
\vspace{-0.2in}
\caption{We find an optimal combination of compatible alliances by solving a weighted max-clique problem. Each node corresponds to an alliance and its weight corresponds to the estimated value of the alliance. There exists an edge between every pair of compatible alliances. We therefore find the combination of alliances with the optimal total value.}
\Description[Weighted Max-Clique Problem]{A weighted, partially connected graph. All cliques whose size can not be increased are encircled. The clique with the highest total weight is marked red.}
\label{fig:alliance_creation}
\end{figure}

Once we have found a good set of alliances, we create the artificial alliance DCs and return the alliances. In practice, we solve this optimization problem using \textsc{TSM-MWC} \cite{jiang2018maxsat}, an exact state-of-the art method for solving the weighted max-clique problem.
An example of such a weighted max-clique problem is given in Figure. \ref{fig:alliance_creation}.

It is possible to increase the efficiency of Algorithm. \ref{alg:create_alliances}, however, since the number of DCs is generally small compared to the number of DOs, the runtime of the matching algorithm will be negligible compared to the model training efforts. For clarity, we show a purely functional version of the algorithm.

\subsection{Ensemble Distillation Procedure}
When a DC participates in one or more alliances, it will train multiple models simultaneously which it has to merge to obtain its global model. In \methodname{}, we use ensemble distillation to merge the different submodels that are created by the alliances and the DC. When we perform ensemble distillation, we want the student model to learn from the joint knowledge of the teacher models. The loss of the student for a given sample is therefore defined based on a measure of distance between the output of the student and the outputs of the teachers.

For our approach, we define the training loss of the student model as in \cite{deng2023hierarchical}. Given a sample $x$, the logits of $m$ teacher models $z_1=\mathcal{T}_1(x), ..., z_m=\mathcal{T}_m(x)$ and the logits of the student model $z_\mathcal{S}=\mathcal{S}(x)$, we define the student loss as follows:

First compute the weights for all teacher logits as:

\begin{displaymath}
w_i^{logits}=\frac{exp(-H(\sigma(z_i))}{\sum_{j=1}^mexp(-H(\sigma(z_j))}
\end{displaymath}

We perform this weighting, because smaller entropy in the output distribution implies that a teacher model is more confident about its prediction, which should therefore be more influential in the training of the student. We then compute the weighted average of the teacher logits:
\begin{displaymath}
    z_\mathcal{T} = \sum_{i=1}^mw_i^{logits}z_i
\end{displaymath}

We calculate the output probability distributions using the softmax function and determine the student loss as a weighted average of the soft KL-Divergence-based loss and the hard cross-entropy loss (see Equation. \eqref{eq:distillation_loss}),

\begin{equation}
\label{eq:distillation_loss}
L(z_\mathcal{S}, Z_\mathcal{T})=\alpha D(P_\mathcal{S} \| P_\mathcal{T}) + (1-\alpha) L_{CE}(Z_\mathcal{T})
\end{equation}

where $Z_\mathcal{T}$ is the set of all teacher logits and the cross-entropy loss $L_{CE}$ is computed based on the assumption that $argmax(P_\mathcal{T}(x))$ is the true label. By minimizing this loss, the student tries to mimic the behavior of the teacher. $\alpha$ is a tunable parameter of the distillation procedure.

\subsection{FL Market using FedCDC}

\begin{figure}[h]
\begin{center}
\includegraphics[width=\linewidth]{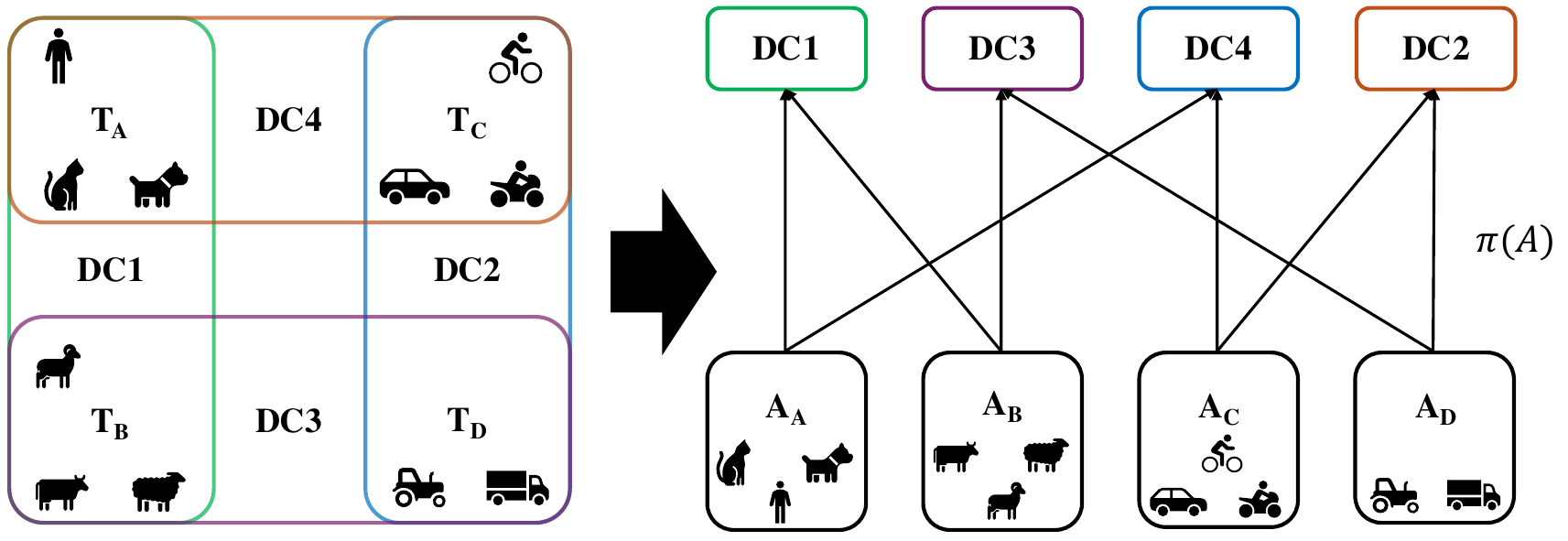}
\end{center}
\vspace{-0.2in}
\caption{A more complex collaboration pattern as it may emerge in an FL Market. 4 Data Consumers want to classify different sets of classes. FedCDC detects 4 separate subtasks and creates corresponding alliances. The alliances train their own models, which are then provided to the alliance participants.}
\label{fig:complex_collaboration}
\Description[Complex Collaboration Pattern]{On the right side of the diagram, a Venn diagram shows the target domains of 4 Data Consumers. Each target domain intersects with two other target domains, such that there are 4 separate intersections. On the right side, 4 alliances, one for each intersection are mapped to the Data Consumers who participate in the alliance. Each alliance is mapped to 2 out of the 4 Data Consumers. The participants of an alliance are the Data Consumers whose target domains form the corresponding intersection in the Venn diagram.}
\end{figure}

In Algorithm. \ref{alg:fl_market_collab}, we describe what integrating FedCDC into a simple FL market would look like. Data Consumers regularly bid for Data Owners in order to recruit them for their training processes. Based on the bids, the DOs are then assigned to the DCs. This matching is determined by an auction mechanism such as a first-price, second-price or VCG auction. It could also be a double auction taking into account bids from DOs. \methodname{} treats this mechanism as a black-box, it only requires the auction to take into account DC-bids, because they are used to detect possible alliances. Since alliances act as synthetic Data Consumers, they are included in this bidding process and recruit their own DOs. However, we do not allow synthetic DCs to join other alliances, so their bidding history is not stored. Afterwards, real and synthetic DCs train their models using an FL framework such as \textsc{FedAvg} or \textsc{FedDF}. The alliance models are then merged with the regular DC models through knowledge distillation. Before the next bidding round, our framework detects new alliance opportunities by using DC metadata, as well as their bidding history during the last $k$ rounds. Alliances are created as described in Algorithm. \ref{alg:create_alliances}. In real FL Markets, more complex collaboration patterns between the DCs may appear, an example is given in Figure. \ref{fig:complex_collaboration}.

\begin{algorithm}[t]
   \caption{Federated Learning Market with FedCDC}
   \label{alg:fl_market_collab}
\begin{algorithmic}
   \STATE {\bf Input:} Data Consumers $\mathcal{C}$, $|\mathcal{C}|=m$; Data Owners $\mathcal{O}$, $|\mathcal{O}|=n$; Bidding history span $k$; Alliance creation parameters $\rho_{AC}=(n_{L, min}, n_{\delta, min})$; Public dataset $\mathcal{D}_{public}$; Distillation parameters $\rho_{ED}=(\alpha,\eta, E)$
   \STATE
   \STATE $r \gets 0$
   \STATE Set of all Alliances: $\mathcal{A} \gets \{\}$
   \STATE Bidding history: $B^{hist} \in \mathbb{R}^{k, m,n}, B^{hist} \gets \mathbf{0}$
   \WHILE{running}
      \STATE Alliance DCs: $\mathcal{C}_A \gets \{C_A | A \in \mathcal{A}\}$
      \STATE Get bids from DCs: $B \in \mathbb{R}^{m+|\mathcal{A}|, n}$
      \STATE Update bidding history using bids from real DCs; $B^{hist,r \mod k} \gets B^{m, n}$
      \STATE Matching based on bids $f: \mathcal{O} \rightarrow \mathcal{C} \cup \mathcal{C}_A$
      \FORALL{$C_i \in \mathcal{C} \cup\mathcal{C}_A$ {\bf in parallel}}
        \STATE $\mathcal{O}(C_i) \gets \{O_j\mid j \in \mathcal{O}: f(j)=i\}$
         \STATE $\textsc{FedAvg}(C_i, \mathcal{O}(C_i))$ (or another FL method)
      \ENDFOR
      \STATE Merge Alliance models with DC models:
      \FORALL{$i \in \mathcal{C}$ {\bf in parallel}}
        \STATE $M^T_i \gets \{M_A|A \in \mathcal{A} : i \in \pi(A)\}$
        \STATE $M^T_i \gets M^T_i \cup \{M'_i\}$
        \FOR{epoch $= 1$ {\bf to} E}
      \FOR{each mini-batch $\mathcal{B_D} \subset \mathcal{D}_{public}$}
         \STATE $Z^i_\mathcal{T} = \{\mathcal{T}^i(\mathcal{B_D}) \mid \mathcal{T}^i \in  M^T_i\}$
         \STATE $z^i_\mathcal{S} = M_i(\mathcal{B_D})$
         \STATE $L = L(z^i_\mathcal{S}, Z_\mathcal{T}^i)$ (Equation. \eqref{eq:distillation_loss})
         \STATE $w_{M_i} \gets w_{M_i} - \eta \nabla_{w_{M_i}} L$
      \ENDFOR
   \ENDFOR
      \ENDFOR
      \STATE $\mathcal{A}' \gets \textsc{CreateAlliances}(\mathcal{C}, \mathcal{O}, B^{hist}, \rho_{AC})$ (Alg. \ref{alg:create_alliances})
      \STATE $\mathcal{A} \gets \mathcal{A} \cup \mathcal{A}' $
      \STATE $r \gets r + 1$
   \ENDWHILE
\end{algorithmic}
\end{algorithm}

\section{Evaluation}
In this section, we first analyze how restricted access to DOs in an FL market affects the training processes of the DCs. Then we test how integrating \methodname{} benefits the DCs in these scenarios. 

\subsection{Experimental Setup}
\paragraph{FL Market}
We model a competitive FL Market as follows:
\begin{itemize}
    \item The market contains 3 Data Consumers who want to perform image classification and 24 Data Owners with labeled image data. 
    \item Each DC $i$ wants to classify a set of $|COI_i|=n_c$ classes. The DCs share half of their classes of interest, i.e. $|COI_A|= |\cap COI_i|=\frac{n_c}{2}$. 
    \item DOs hold data for classes in $COI_A$ or for classes in $COI_i\setminus COI_A$ for some $i$. This means that either all DCs want to recruit a DO or exactly one DC wants to recruit it, which is the optimal setting for our collaboration approach. We therefore have $n_{DC}+1=4$ groups of DOs. There are 6 DOs in every group. Inside every group, the data is partitioned homogeneously between the 6 DOs.
    \item Since all DCs try to recruit DOs with data from $COI_A$, we randomly partition the DOs between the DCs, such that every DC is assigned 2 shared DOs. This simulates a scenario in which all DCs have similar budgets, so they have similar bidding power. The remaining DOs are assigned to the unique DC who is interested in them. We perform DC-DO-matching every 5 communication rounds.
\end{itemize}
    
\paragraph{Scenarios}
We use the market model as well as the DC strategies described above to simulate a competitive FL market in which DCs have limited access to DOs. In order to evaluate how this limitation affects the training processes of the DCs, we compare this to the ideal setting in which every DC can always recruit every DO. We then test how integrating \methodname{} into the competitive market benefits the performance of the DC models.
\begin{itemize}
    \item \textbf{Unrestricted DO-Access}: Each DC can recruit every DO in every FL round. This is an ideal baseline which does not represent the actual scenario in which \methodname{} is applied. The goal of FedCDC is to achieve similar performances to this baseline.
    \item \textbf{Restricted DO-Access}: The DCs can only recruit a part of the DOs they want in every FL round, just like it would happen in a real FL Market. This is implemented using the market model described above.
    \item \textbf{\methodname{}}: The FL Market runs in the competitive setting for 10 rounds. Then, the DCs form an alliance using \methodname{}. A separate alliance DC is created, trained on the DOs in $COI_A$ and merged into the DC models after every round.
\end{itemize}

\paragraph{Aggregation Methods}
We perform our tests for \textsc{FedAvg}\cite{mcmahan2017communication} and \textsc{FedDF}\cite{lin2020ensemble}. The DCs always aggregate the local models of all DOs that they were assigned. In the scenarios without collaboration, the aggregation method is used to create the global DC model from the locally trained models as in standard FL. In the \methodname{} scenario, the alliance DC, as well as the DC-experts use the aggregation method to create intermediary models, which are then merged using ensemble distillation. 

\paragraph{Datasets and Data Heterogeneity}
We perform our tests on the FMNIST\cite{xiao2017fashion}, CIFAR-10\cite{giuste2020cifar} and CIFAR-100\cite{zheng2024comparative} datasets. We split the training set into a public dataset, a validation set for every DC and a training set for every DO. The public dataset, which is used for knowledge distillation, contains 5000 unlabeled and homogeneously distributed samples. The validation sets for each DC contain 2000 samples, which are evenly divided into the $n_c$ classes of interest of the DC. Each DO owns a training set of 1000 samples which is evenly divided into the $\frac{n_c}{2}$ classes determined by the group that the DO belongs to. For CIFAR-10 and FMNIST, which consist of 10 different classes, we set $n_c=4$, for CIFAR-100, which consists of 100 different classes, we set it to $n_c=40$.

\paragraph{Parameters and Models} We train the models for $N=50$ FL rounds. For local training as well as ensemble distillation, we use the Adam Optimizer with learning rate $0.001$ and batch size $32$. For \textsc{FedAvg}, we train the local models for $5$ epochs. For \textsc{FedDF} we train the local models for $10$ epochs and perform distillation for $5$ epochs. In FedCDC, we perform ensemble distillation for $10$ epochs. For our distillation procedure, we set $\alpha=1$, so we only use the soft-label-loss. All models are of the ResNet-18 architecture.

\paragraph{Evaluation Metrics} 
\begin{itemize}
    \item \textbf{Accuracy:} For each scenario, we report the average test accuracy of the final DC models after $N$ rounds of training. We evaluate the test accuracies on the models with the best validation accuracies.
    \item \textbf{Convergence:} We also provide plots that detail the average accuracy of the DC models after every FL round to investigate how quickly the models converge.
\end{itemize}

\subsection{Results}

\begin{figure}[h]
\begin{center}
\includegraphics[width=\linewidth]{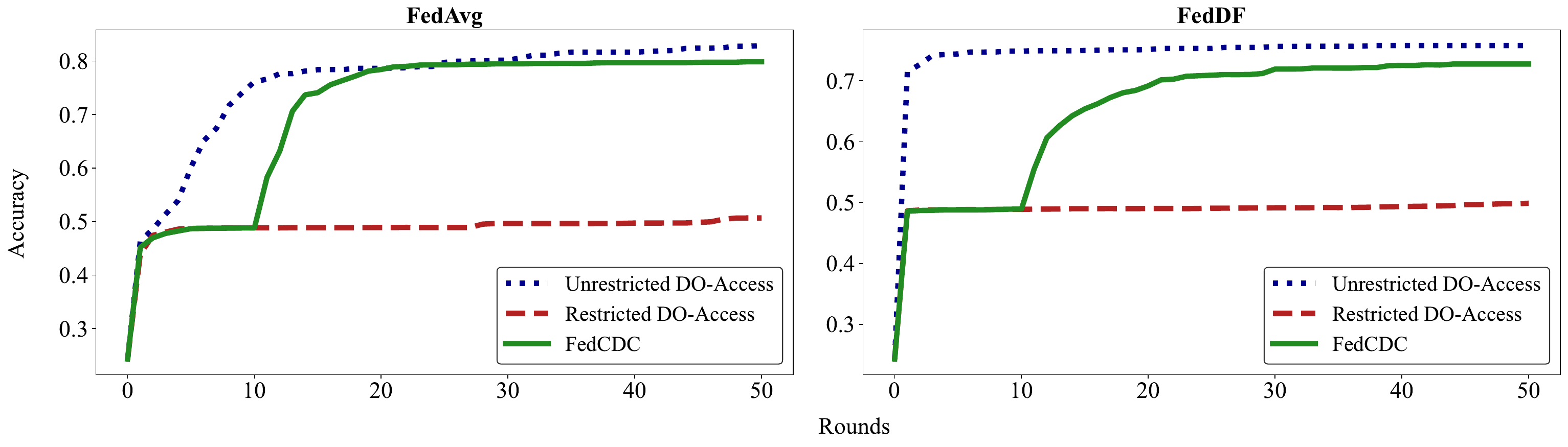}
\end{center}
\vspace{-0.2in}
\caption{Average test accuracy of all DCs after every communication round for the different scenarios. The impact of restricted DC-Access, as well as collaboration are examined for FMNIST and FedAvg/FedDF.}
\label{fig:results_fmnist}
\Description{Line graphs for the 3 scenarios when FedAvg and FedDF are used for aggregation. In the scenario with restricted access to Data Owners, the accuracy quickly converges to its final value. When the access is unrestricted, the accuracy keeps increasing for more rounds, until it reaches its final value. In the FedCDC scenario, when the Data Consumers start collaborating, the accuracy starts increasing until it converges to a value similar to the baseline with unrestricted access. The final accuracies are given in Table 1.}
\end{figure}

In our experiments (Table. \ref{tab:results}) we observe that restricted access to DOs negatively impacts DC training processes. Since for each DC, 6 unique DOs, but only 2 shared DOs are aggregated every round, the model converges to a suboptimal equilibrium in which it is biased towards the classes held by the unique DOs. When applying FedCDC, we consistently improve the average accuracy of the DCs. Since the DCs share access to the DOs and can therefore use all of the 6 shared DOs in every round, the bias is avoided. When the DCs start collaborating, their average accuracy converges to a better value due to the more balanced access to data.

\begin{table*}[!h]
\caption{The final average accuracies of the 3 DC models in the different experimental settings.}
\label{tab:results}
\begin{tabular}{cccc}
\toprule
 & Unrestricted DO-Access        & Restricted DO-Access         & FedCDC          \\ \midrule
FMNIST, FedAvg      & $82.90 \pm 0.30$        & $50.67 \pm 0.10$         & $79.88 \pm 0.15$          \\
FMNIST, FedDF       & $75.80 \pm 0.16$        & $49.89 \pm 0.02$         & $72.75 \pm 0.30$         \\
CIFAR-10, FedAvg    & $66.91 \pm 0.07$        & $45.82 \pm 0.02$         & $68.54 \pm 0.05$          \\
CIFAR-10, FedDF     & $67.97 \pm 0.11$        & $45.53 \pm 0.03$         & $63.03 \pm 0.16$          \\
CIFAR-100, FedAvg   & $45.99 \pm 0.02$        & $28.59 \pm 0.01$         & $38.32 \pm 0.01$          \\
CIFAR-100, FedDF    & $39.38 \pm 0.02$        & $25.94 \pm 0.01$         & $32.40 \pm 0.01$         \\ \bottomrule
\end{tabular}
\end{table*}

\begin{figure}[h]
\begin{center}
\includegraphics[width=\linewidth]{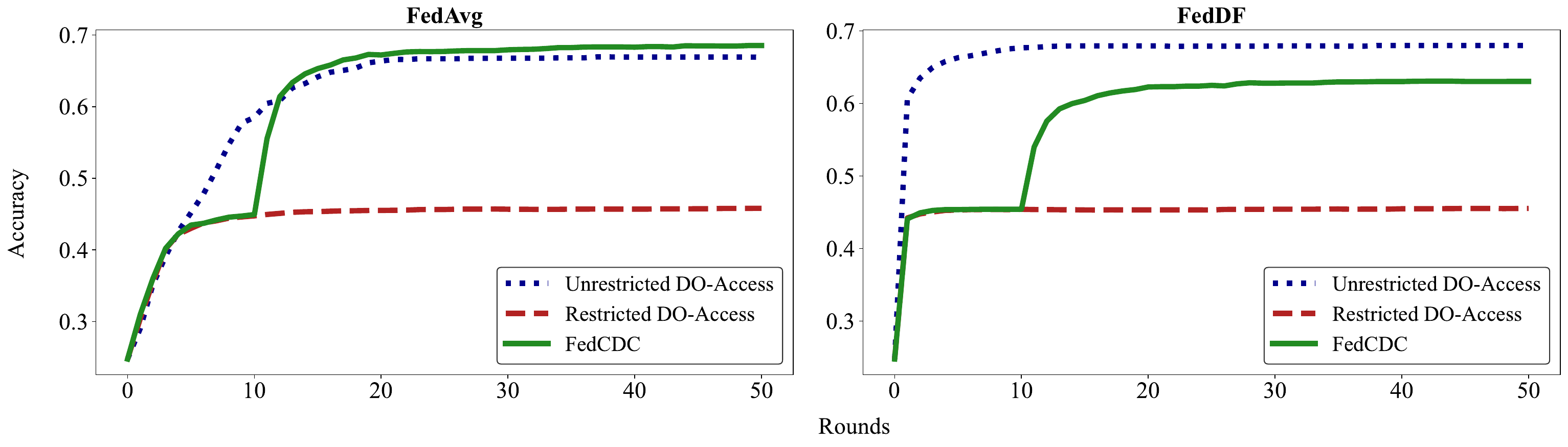}
\end{center}
\vspace{-0.2in}
\caption{Average test accuracy of all DCs after every communication round for the different scenarios. The impact of restricted DC-Access, as well as collaboration are examined for CIFAR-10 and FedAvg/FedDF.}
\label{fig:results_cifar10}
\Description{Line graphs for the 3 scenarios when FedAvg and FedDF are used for aggregation. In the scenario with restricted access to Data Owners, the accuracy quickly converges to its final value. When the access is unrestricted, the accuracy keeps increasing for more rounds, until it reaches its final value. In the FedCDC scenario, when the Data Consumers start collaborating, the accuracy starts increasing until it converges to a value similar to the baseline with unrestricted access. When using FedDF instead of FedAvg, the final accuracy in the collaboration scenario decreases. The final accuracies are given in Table 1.}
\end{figure}

\begin{figure}[h]
\begin{center}
\includegraphics[width=\linewidth]{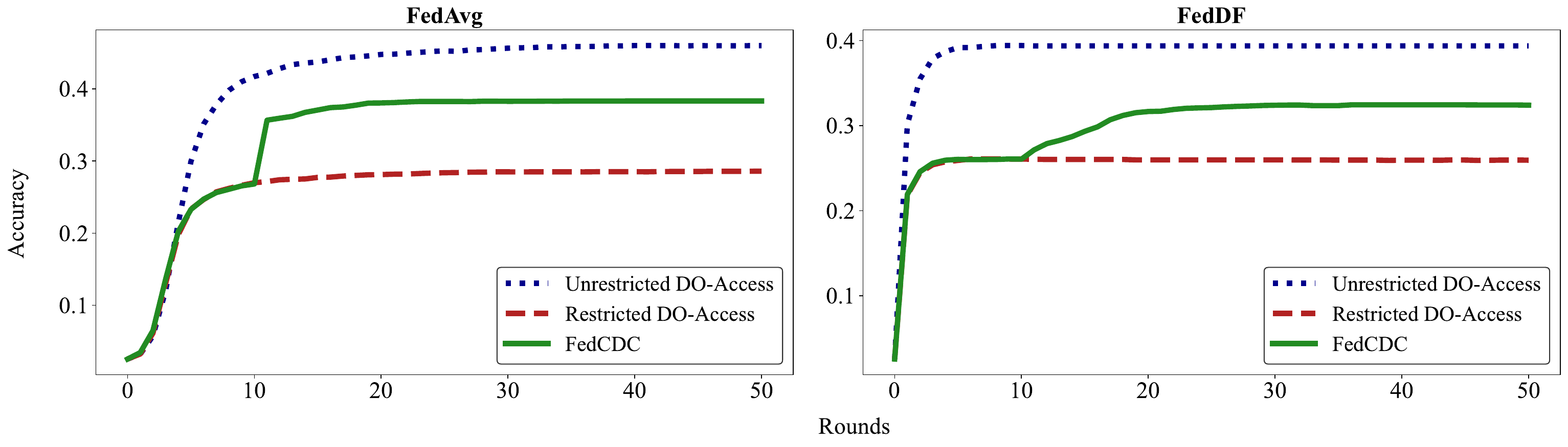}
\end{center}
\vspace{-0.2in}
\caption{Average test accuracy of all DCs after every communication round for the different scenarios. The impact of restricted DC-Access, as well as collaboration are examined for CIFAR-100 and FedAvg/FedDF.}
\label{fig:results_cifar100}
\Description{Line graphs for the 3 scenarios when FedAvg and FedDF are used for aggregation. In the scenario with restricted access to Data Owners, the accuracy quickly converges to its final value. When the access is unrestricted, the accuracy keeps increasing for more rounds, until it reaches its final value. In the FedCDC scenario, when the Data Consumers start collaborating, the accuracy starts increasing until it converges to a value between the other two scenarios. When using FedDF instead of FedAvg, the final accuracy in the collaboration scenario, as well as the baseline with unrestricted access, decrease. The final accuracies are given in Table 1.}
\end{figure}

The exact relation between the accuracies in the unrestricted and the FedCDC scenarios depends on the aggregation method that is applied, as well as the dataset. This is because from the perspective of a single DC, our framework can be seen as an alternative aggregation method similar to \textsc{FedHKT}. Depending on the exact settings, different aggregation methods may perform better or worse than others. Furthermore, FedCDC works better for FMNIST (Figure. \ref{fig:results_fmnist}) and CIFAR-10 (Figure. \ref{fig:results_cifar10}) than for CIFAR-100 (Figure. \ref{fig:results_cifar100}). In our framework, while the alliance model is trained on the intersection of the participant tasks, its output logits represent the union of the participant tasks. We choose this design, because it simplifies the merging procedure and because it improves the flexibility of the alliance model if a shared DO holds data that not all participants are interested in. However, in our experiments for CIFAR-100, this may negatively impact the distillation procedure due to the higher number of classes.

\subsection{Discussion}

In our experiments, we observe that reduced access to Data Owners can severely limit the performance of DC models. Similar to conventional machine learning, where you need a diverse and well-balanced dataset to train your model, FL tasks need access to a diverse and well-balanced set of Data Owners to achieve optimal model performance. In real FL Markets, the access to Data Owners may be limited by competition between DCs for DOs with limited computational resources, as well as DC budgets, which may be too small to recruit a diverse set of DOs. In both cases, this problem can be solved by applying our collaboration approach. Since the DCs train a shared model together, there is less competition for the DOs and the effective budget of the DCs is bigger. In particular, it follows from Equation. \eqref{eq:alliance_budget} that the effective budget of $C_i$ who participates in an alliance $A$ is increased by the sum of payments of all other participants (Equation. \eqref{eq:effective_budget}).

\begin{equation}
    \label{eq:effective_budget}
        b'(C_i)=b(C_i)+\sum_{C_j\in \pi(A)\setminus C_i}payment(C_j \rightarrow A)
\end{equation}

Therefore, while this paper does not focus on budget allocation, it is clear that collaboration has significant advantages regarding budgets and costs. We note that in order to ensure fairness, the alliance budget either has to be split up uniformly between participants (Equation. \eqref{eq:equal_payments}) 

\begin{equation}
    \label{eq:equal_payments}
        \forall i,j\in \pi(A): payment(C_i \rightarrow A)=payment(C_j \rightarrow A)
\end{equation}

or the reward that a participant gains from an alliance has to depend on its contribution. Since a participant's reward depends on the quality of the model that it receives from the alliance, the latter approach may be realized by sending simpler models to participants with smaller contributions, similar to \cite{wang2024fedsac} in traditional Federated Learning.

In our experiments, the DCs always aggregate all of the local models. By using client selection mechanisms, the impact of restricted DO-access may be reduced. Approaches which increase the "smartness" DC agents by using mechanisms such as client selection and contribution evaluation are orthogonal to \methodname{} and their interplay with our approach, especially in settings with more complex data partitioning, can be studied in future work.

\methodname{} does not allow recursive alliances, an alliance DC can not join another alliance. This simplifies the optimization process of the alliance creation procedure, as well as the coordination between participants in an alliance. However, it may be possible to improve the collaboration framework further by allowing recursive alliances, which would lead to more flexible and fine-grained collaboration patterns. Furthermore, we restrict the preferences that DCs can express for alliances and combinations of alliances to boolean choices. By allowing more detailed preferences, e.g. preference lists, and modeling the optimization problem as a hedonic game, we may find solutions that reflect the true preferences of the DCs more accurately.

\section{Conclusion}
In this paper, we introduced \methodname{}, a framework enabling Data Consumers (DCs) to collaborate within competitive FL markets. By jointly training a submodel and transferring its knowledge to each DC’s model via ensemble distillation, our approach allows DCs to leverage Data Owners (DOs) more effectively and significantly enhance their model performance. Furthermore, our alliance-creation mechanism supports the formation of complex collaboration patterns, thereby increasing overall market efficiency.

Although our work primarily targets image classification tasks, the underlying concept of alliances can be extended to any modular task—i.e., one that can be decomposed into subtasks. This framework raises new challenges related to the design of DC agents and budget allocation strategies. We leave the exploration of fair, incentive-compatible, and socially optimal solutions to these challenges for future research. Additionally, further refinement of the ensemble distillation process could not only improve model performance but also potentially eliminate the need for a public dataset.

Finally, while Federated Learning inherently leverages data modularity by facilitating collaborative training on distributed datasets, our proposed framework expands this notion to include task modularity. This broader perspective offers a promising avenue for advancing cooperative techniques in FL markets.

\clearpage
\bibliographystyle{ACM-Reference-Format}
\bibliography{main}

\end{document}